\documentclass{Interspeech2024}

\interspeechcameraready 

\usepackage{times,latexsym}
\usepackage{url}
\usepackage[T1]{fontenc}
\usepackage{microtype}
\usepackage{booktabs}
\usepackage{multirow}
\usepackage{enumitem}
\usepackage{xurl}

\usepackage{graphicx}
\graphicspath{ {./images/} }

\usepackage{xspace,mfirstuc,tabulary}
\usepackage{tablefootnote}

\title{Performant ASR Models for Medical Entities in Accented Speech}

\name[affiliation={ { \dagger }1,6,*}]{Tejumade}{Afonja}
\name[affiliation={ { \dagger }2,*}]{Tobi}{Olatunji}
\name[affiliation={3,*}]{Sewade}{Ogun}
\name[affiliation={4,*}]{\\Naome A.}{Etori}
\name[affiliation={2,*}]{Abraham}{Owodunni} 
\name[affiliation={5,*}]{Moshood}{Yekini}

\address{\small
  $^1$CISPA Helmholtz Center for Information Security
  $^2$Intron Health
  $^3$Université de Lorraine, CNRS, Inria, LORIA, F-54000 Nancy, France
  $^4$University of Minnesota - Twin Cities, USA
  $^5$African Masters of Machine Intelligence, AIMS/AMMI\\
  $^6$AI Saturdays Lagos
  $^*$Masakhane NLP
  \thanks{ $\dagger$Equal contribution.}
  }
\email{tejumade.afonja@cispa.de, tobi@intron.io}


\begin{document}

\maketitle

\newcommand{\medication}{\texttt{MED}}
\newcommand{\condition}{\texttt{COND}}
\newcommand{\anatomy}{\texttt{ANA}}
\newcommand{\phinfo}{\texttt{PHI}}
\newcommand{\ttp}{\texttt{TTP}}

\begin{abstract}

Recent strides in automatic speech recognition (ASR) have accelerated their application in the medical domain where their performance on accented medical named entities (NE) such as drug names, diagnoses, and lab results, is largely unknown. We rigorously evaluate multiple ASR models on a clinical English dataset of 93 African accents. Our analysis reveals that despite some models achieving low overall word error rates (WER), errors in clinical entities are higher, potentially posing substantial risks to patient safety. To empirically demonstrate this, we extract clinical entities from transcripts, develop a novel algorithm to align ASR predictions with these entities, and compute medical NE Recall, medical WER, and character error rate. Our results show that fine-tuning on accented clinical speech improves medical WER by a wide margin (25-34~\% relative), improving their practical applicability in healthcare environments.
\end{abstract}

\section{Introduction}
\label{sec:introduction}
In recent years, significant advances have been made in accented speech recognition with state-of-the-art (SOTA) automatic speech recognition (ASR) models proficiently transcribing diverse linguistic interactions \cite{radford2022robust, 9585401, Baevski2020wav2vec2A}. However, the effectiveness of these models in clinical or medical settings,\footnote{We use the terms `clinical' and `medical' interchangeably to encompass all aspects related to the practice of medicine and patient care.} where nuanced communication is paramount, remains a challenge \cite{olatunji2023afrispeech}. This becomes particularly evident when clinicians with non-western accents document critical medical information using ASR technology. While these SOTA models achieve low word error rates (WER) on general speech, they commonly struggle with accurately transcribing clinical named entities (NE), e.g., see Table~\ref{tab:medtextalign}. The domain-specific nature of clinical documentation introduces a vulnerability that could have severe consequences for patient well-being -- minor inaccuracies in essential elements like drug names, diagnoses, lab results, and lesion measurements (for example, writing \textit{renal} instead of \textit{adrenal}, or \textit{hyper-} instead of \textit{hypo-}) could potentially risk patient safety and expose clinicians to avoidable litigation \cite{ajami2016use}.
To empirically expose this problem, we analyze the performance of several SOTA open-source and commercial ASR models on medical NEs (MNEs). Our investigation reveals that current SOTA general-purpose multilingual ASR models while excelling in cross-domain scenarios, exhibit sub-optimal Recall rates for MNEs in accented speech. This limitation diminishes the practical utility of these models in healthcare settings, underscoring the need for specialized solutions. 

Our contributions are as follows:
\begin{enumerate}
    \item We benchmark 19 open-source and commercial ASR models on African accented clinical speech highlighting the deficiencies of existing architectures in accurately recognizing accented MNEs.
    \item We introduce metrics for evaluating medical NER performance in the context of accented speech, including medical named entity Recall, medical WER (M-WER), and medical character error rate (M-CER).
    \item We develop a novel fuzzy string matching algorithm to better align ASR-predicted noisy NEs to ground truth NEs for more nuanced analysis. 
    \item We demonstrate that supervised fine-tuning substantially enhances accented medical NER, making ASR models more applicable and reliable in real-world clinical scenarios.
\end{enumerate}

\begin{table*}[ht!]
\caption{Predicted sentences from selected ASR models compared to the reference sentence. 
}
\label{tab:medtextalign}
\small
\centering
\begin{tabular}{l|l}
\hline
Model & Sentence\\
\hline 
Reference sentence & \textbf{lungs clear} but dim scattered \textbf{rhonchi} nonproductive \textbf{cough}.\\
\hline
Xlsr-53-en &  \underline{longscler} bout deim scattered \underline{rong i} non-productive hol\\
Whisper-medium & \underline{non-scler,} but dim-scattered \underline{ronchi}, non-productive hub. \\
\hline
GCP [Medical] & \underline{lungs , clear} . budan scattered \textbf{rhonchi} . nonproductive\\
AWS [Medical] (Primary Care) & \underline{last clear} but deems scattered \textbf{rhonchi} nonproductive.\\
Whisper-medium-clinical	&  \textbf{lungs clear} but dim scattered \textbf{rhonchi} nonproductive \textbf{cough}. \\
\hline
\end{tabular}

\vspace{-15pt}
\end{table*}

\section{Related Work}
\label{sec:related}
Recently, authors in~\cite{olatunji2023afrinames} highlighted the challenges faced by popular ASR models in recognizing African named entities like persons, locations, and organizations from accented speech. 
They improved entity WER through data augmentation techniques. 
In the medical domain, the authors in \cite{adedeji2024sound} relied on large language models for correcting medical ASR transcription errors. The work of \cite{jiang2021sequence} also used a sequence-to-sequence model to correct clinical ASR errors.  
Accurately detecting and classifying medical named entities from text has been explored in \cite{bhatia2019comprehend, zhou2021clinical} where \cite{bhatia2019comprehend} identified five key MNEs, and employed deep learning and multi-task learning approaches to extract crucial information from clinical narratives. Also, the authors in \cite{zhou2021clinical} developed an ensemble of deep contextual models trained on clinical corpora from PubMed to enhance clinical NE recognition. The work of \cite{suominen2015benchmarking} separately benchmarked clinical speech recognition and entity extraction. Additionally, a production-scalable BiLSTM-CNN-Char framework with pre-trained embedding was designed by \cite{kocaman2022accurate}, which was shown to achieve better performance in speed and prediction compared to the SOTA models and commercial clinical NE recognition solutions. The authors in \cite{hu2024improving} proposed a clinical task-specific prompting framework that adopts entity definitions, annotation guidelines and samples, and error analysis-based instructions.  
However, research benchmarking SOTA ASR models on accented medical NE transcription or recognition is still lacking.

\section{Approach}
\begin{figure}[t]
  \centering
  \includegraphics[width=\linewidth]{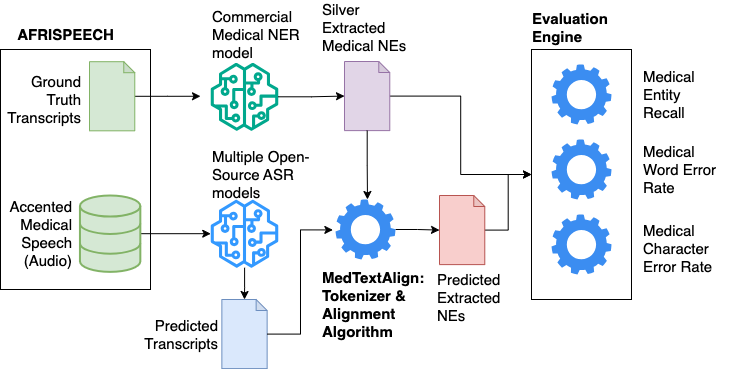}
  \caption{Methodology: Ground truth transcripts are passed to a commercial medical NER model, and audios are passed through multiple ASR models. Predicted medical entities are extracted using the MedTextAlign algorithm. Metrics are computed over silver NEs}
  \label{fig:approach}
  \vspace{-10pt}
\end{figure}

We investigated this problem by evaluating 19 open-source and commercial ASR systems on a dataset of
African-accented clinical speech. A schematic is shown in Figure~\ref{fig:approach}. The dataset, medical NE extraction approach, ASR models, and evaluation methods are described below. 

\subsection{Data}
For this analysis, we leveraged AfriSpeech-200 \cite{olatunji2023afrispeech}, a 200-hour Pan-African accented English speech corpus for clinical and general domain ASR with 120 accents, and over 2,300 unique speakers from over 10 African countries. The dataset statistics and NE categories are shown in Table~\ref{tab:splits}. Our evaluation focuses on the clinical domain test subset. After filtering out texts lacking sufficient medical context, we retained a total of 2,844 samples, encompassing 93 different accents.

\vspace{-5pt}
\begin{table}[ht!]
\caption{Dataset splits showing the number of speakers, the number of clips, speech duration, and medical named entity category counts in train/dev/test splits.}
\label{tab:splits}
\addtolength{\tabcolsep}{-0.55pt}
\centering
\begin{tabular}{l|c|c|c}
\hline
Item & Train & Dev  & Test\\
\hline
Number of speakers & 1466 & 247 & 750 \\
Duration (in hours) & 173.4 & 8.74 & 18.77 \\
Number of accents & 71 & 45 & 108 \\
Number of clips/speaker & 39.56 & 13.08 & 8.46 \\
Number of speakers/accent & 20.65 & 5.49 & 6.94 \\
\hline

\#~clinical domain clips (61.80~\%) & 36318 & 1824 & 3623 \\
\#~general domain clips (38.20~\%) & 21682 & 1407 & 2723 \\
\hline
\multicolumn{4}{c}{Medical named entity category count}\\
\hline
Medication (\medication)  & 4164 & 132 & 276 \\
Medical condition (\condition) & 18804 & 901 & 1414\\
Anatomy (\anatomy) & 13650 & 645 & 927\\
Test treatment procedure (\ttp) & 10713 & 428 & 893 \\
Protected health information (\phinfo) & 3449 & 105 & 253\\
\hline
\end{tabular}
\vspace{-15pt}
\end{table}

\begin{table*}[!h]
\caption{Output from the Wavlm-libri-clean-100h-base model, entities matched exactly are highlighted in bold, while near matches identified through our MedTextAlign strategy are underlined.}
\label{tab:medtextalign_wavlm}
\centering
\begin{tabular}{p{1\columnwidth}|p{1\columnwidth}}
\hline
Reference  & Prediction \\
\hline 
unlike \textbf{quinidine}, \textbf{disopyramide} does not increase the plasma concentration of \textbf{digoxin} in patients & anlike \underline{quinidan}, \underline{disopiramid} dos not incruse the plasma concentration of \underline{dikod sin} in pesion\\
\hline
except for \textbf{ketamine}, the following agents have no \textbf{analgesic properties} and do not cause \textbf{paralysis} or \textbf{muscle relaxation} & except for \underline{ketami}, befullin agents have no \underline{anagesic propatis} and do not cose \textbf{paralysis} o \underline{mozul relaxition} \\

\bottomrule
\end{tabular}

\end{table*}

\begin{table*}[!ht]
    \caption{Performance evaluation of benchmarked models on AfriSpeech-200 clinical domain test dataset. We report WER comparing transcript and prediction, alongside specific metrics including the medical WER (M-WER), the medical CER (M-CER), and the Recall for the different entities, medication (\medication), anatomy (\anatomy), medical condition (\condition), test treatment procedure (\ttp), and protected health information (\phinfo).}
    \label{tab:models_benchmarks}
    \centering
    \small
    \begin{tabular}{lc|cc|ccccc}
\hline
\multicolumn{2}{r|}{} & \multicolumn{2}{r|}{} & \multicolumn{5}{c}{Recall($\uparrow$)} \\
Models  &    WER & M-WER & M-CER & \medication & \anatomy & \condition &   \ttp &  \phinfo \\
\hline
Pretrained \\
\hline
Wavlm-libri-clean-100h-base            &  0.902 &  0.944 &  0.504 &  0.019 &  0.069 &  0.063 &  0.056 &  0.034 \\
Wavlm-libri-clean-100h-large          &  0.784 &  0.852 &  0.307 &  0.029 &  0.177 &  0.142 &  0.110 &  0.043\\
Hubert-large-ls960-ft                  &  0.712 &  0.758 &  0.279 &  0.070 &  0.282 &  0.258 &  0.168 &  0.069 \\
Hubert-xlarge-ls960-ft                &  0.722 &  0.770 &  0.275 &  0.067 &  0.284 &  0.262 &  0.166 &  0.075 \\
Wav2vec2-large-robust-ft-swbd-300h     &  0.907 &  0.919 &  0.367 &  0.040 &  0.129 &  0.139 &  0.112 &  0.051 \\
Wav2vec2-large-960h                     &  0.796 &  0.846 &  0.345 &  0.032 &  0.189 &  0.171 &  0.109 &  0.049 \\
Wav2vec2-large-960h-lv60-self          &  0.694 &  0.753 &  0.277 &  0.064 &  0.309 &  0.254 &  0.173 &  0.087 \\
Wav2vec2-xls-r-1b-english             &  0.666 &  0.729 &  0.266 &  0.081 &  0.251 &  0.249 &  0.227 &  0.138 \\
Wav2vec2-large-xlsr-53-english        &  0.646 &  0.710 &  0.272 &  0.072 &  0.261 &  0.256 &  0.201 &  0.095 \\
Whisper-small-en                        &  0.486 &  0.566 &  0.225 &  0.215 &  0.571 &  0.536 &  0.475 &  0.300 \\
Whisper-small                          &  0.451 &  0.567 &  0.216 &  0.235 &  0.566 &  0.541 &  0.486 &  0.301 \\
Whisper-medium-en                       &  0.415 &  0.504 &  0.188 &  0.330 &  0.636 &  0.601 &  0.532 &  0.300 \\
Whisper-medium                          &  0.392 &  0.487 &  0.174 &  0.343 &  0.680 &  0.627 &  0.568 &  0.335 \\
Whisper-large                          &  0.373 &  0.454 &  0.154 &  0.425 &  0.717 &  0.667 &  0.597 &  0.331 \\
\hline
Commercial \\
\hline
Azure                                         &  0.442 &  0.491 &  0.216 &  0.611 &  0.660 &  0.623 &  0.515 &  0.261 \\
AWS                                             &  0.540 &  0.660 &  0.249 &  0.212 &  0.523 &  0.485 &  0.382 &  0.246 \\
AWS [Medical] (Primary Care)                     &  0.516 &  0.553 &  0.218 &  0.572 &  0.644 &  0.567 &  0.494 &  0.204 \\
GCP                                            &  0.622 &  0.634 &  0.391 &  0.386 &  0.425 &  0.380 &  0.332 &  0.177 \\
GCP [Medical]                                   &  0.527 &  0.434 &  0.211 &  0.568 &  0.701 &  0.565 &  0.513 &  0.184 \\
\hline
Fine-tuned on AfriSpeech-200 \\
\hline
Wav2vec2-large-xlsr-53-english-general  &  0.473 &  0.680 &  0.235 &  0.144 &  0.294 &  0.300 &  0.297 &  0.300 \\
Wav2vec2-large-xlsr-53-english-both      &  0.308 &  0.467 &  0.135 &  0.451 &  0.658 &  0.576 &  0.567 &  0.356 \\
Wav2vec2-large-xlsr-53-english-clinical &  0.307 &  0.465 &  0.133 &  0.496 &  0.689 &  0.584 &  0.588 &  0.291 \\
Whisper-medium-general                    &  0.532 &  0.711 &  0.347 &  0.114 &  0.314 &  0.279 &  0.282 &  0.325 \\
Whisper-medium-clinical                   &  0.264 &  0.388 &  0.136 &  0.659 &  0.806 &  0.712 &  0.706 &  0.405 \\
Whisper-medium-both                        &  \textbf{0.241} &  \textbf{0.365} &  \textbf{0.118} &  \textbf{0.731} &  \textbf{0.822} &  \textbf{0.725} &  \textbf{0.726} &  \textbf{0.490} \\
\hline
\end{tabular}
\end{table*}

\subsection{ASR Models}
We evaluated several open-source and commercial (general-purpose and medical) ASR systems covering multiple SOTA ASR architectures shown in Table~\ref{tab:models_benchmarks}. 
In addition, we selected two models for fine-tuning based on the model performance reported in \cite{olatunji2023afrispeech} and our computational constraints.

\subsection{Named Entity Extraction}

\subsubsection{Extracting Ground Truth Entities}
Since the dataset was not annotated with MNEs, we leveraged a commercially available medical NER model, Amazon Comprehend Medical \cite{bhatia2019comprehend},\footnote{Amazon Comprehend Medical at \url{https://aws.amazon.com/comprehend/medical/}} to automatically extract medical NEs from ground truth transcripts. This service has been publicly benchmarked against other NER systems by GigaOm\footnote{GigaOm Clinical NLP Benchmark at \url{https://gigaom.com/report/healthcare-natural-language-processing/}} and \cite{kocaman2022accurate}, and has good accuracy in predicting
multiple medical NE categories. We call these silver annotations as these are not human annotations.

\subsubsection{Selected Named Entities}
We focused on five key medical named entity categories: medication (\medication), medical condition (\condition), anatomy (\anatomy), test treatment procedure (\ttp), and protected health information (\phinfo). These categories cover a wide range of entities, including medication names, dosages, diagnoses, signs, symptoms, and protected health information such as names, addresses, ID numbers, etc. The distribution of entities across these categories for each dataset split is detailed in Table~\ref{tab:splits}.

\subsubsection{MedTextAlign: Extracting Predicted Named Entities}
To evaluate predicted MNEs, a naive method is to use an NER model to identify the ASR-predicted MNEs. However, given that the ASR predictions are noisy, often having different lengths and spellings than ground-truth NEs, and single-word to multi-word entity mismatches exist, these issues pose a challenge for most NER models, making them inadequate. For example, ``analgesic properties'' is misspelled as ``anagesic propatis'', ``digoxin'' wrongly transcribed as ``dikod sin'', and ``spironolactone'' as ``spiro no lactone''. An alignment algorithm was thus needed to better match ground-truth MNEs. Therefore, we developed MedTextAlign, a  
solution that 
uses a fuzzy string-matching algorithm to better align the predicted to the ground truth (silver) MNEs. 

MedTextAlign first tokenizes the predicted transcript, creates a candidate list of unigrams, bigrams, and trigrams from the predicted transcript, then leverages a fuzzy string matching algorithm\footnote{We used the python SequenceMatcher ratio() method at \url{https://docs.python.org/3/library/difflib.html} and set the cut-off threshold to 0.5.} to compare with each MNE from the ground truth transcript to find the closest match. The fuzzy match, akin to measuring the longest common character subsequence (e.g., in ROUGE-L \cite{lin-2004-rouge}), produces a score between 0 and 1 for each string pair, with 1 indicating a perfect match, enabling effective matching of nearly correct spellings in the transcript, e.g., matching wrongly spelled ``quinidan'' or ``disopiramid'' (see Table~\ref{tab:medtextalign_wavlm}).

Although not perfect, this strategy proved to be very effective. In Table~\ref{tab:medtextalign_wavlm}, we underline approximate entity matches and put in bold-face exact matches given by the Wavlm-libri-clean-100h-base model.

\subsection{Evaluation Metrics}
Given the challenges with ASR alignment, conventional ASR or NER metrics fail to effectively measure the model's ability to transcribe medical entities. Consequently, to comprehensively assess the performance of these ASR models, we opted for a broad range of metrics that cover various evaluation dimensions.

\begin{enumerate}
    \item Recall: An information retrieval metric that computes the proportion of recovered correct (exact match) entities in the prediction. Precision and F1 score were not computed because they are overly sensitive to ASR noise or errors. Higher Recall is better.
    \item Word error rate (WER): a word-level metric that evaluates insertions, deletions, and substitutions in the predicted sequence. Lower is better. 
    \item Medical WER (M-WER): WER computed between the ground truth MNEs and their aligned MNEs in the prediction alone. This isolates WER on MNEs of interest while ignoring all other words. All ground truth MNEs in each sample are concatenated with intervening spaces. The MNEs recovered by MedTextAlign are also concatenated in the same way. WER is then calculated between resulting sequences. Lower is better. 
    \item Medical CER (M-CER): Similar to medical WER, but at the character level. M-CER measures the severity of ASR misspellings. Lower is better. 

\end{enumerate}

\section{Experiments}

\subsection{Benchmarking}
We compared SOTA open-source pre-trained ASR models: Whisper \cite{radford2022robust}, Wav2vec2 \cite{Baevski2020wav2vec2A}, XLSR \cite{Babu2022XLSRSC}, Hubert \cite{9585401}, WavLM \cite{chen2022wavlm}, alongside commercial clinical and non-clinical ASR systems; Azure~\cite{azure}, AWS~\cite{aws}, and GCP~\cite{gcp}. For all open-source pre-trained ASR models, we refer readers to read their respective papers for details on pretraining corpora, model architecture, and hyperparameters. In addition, we used the Hugging Face transformer library \cite{wolf-etal-2020-transformers} for inference. For each model, we show results on the AfriSpeech clinical domain test set in Table~\ref{tab:models_benchmarks}. 

\subsection{Fine-tuning}

For the fine-tuning experiments,
we fine-tuned 
the ASR models on three domains: (1) \emph{general} domain (21,682 clips), (2) \emph{clinical} domain (36,318 clips), and (3) \emph{both} domains (58,000 clips). We fine-tuned the models using each domain's training set and tested on the clinical domain test set to investigate the effect of out-of-domain accented data on model performance. Additionally, based on the benchmark results in Table~\ref{tab:models_benchmarks} and GPU memory constraints, two top performing open-source model architectures, Whisper-medium \cite{radford2022robust} and Wav2vec-large-xlsr-53 (XLSR-53) \cite{grosman2021xlsr53-large-english}, were selected for fine-tuning.

XLSR-53 (378.9~M parameters) is an encoder-decoder architecture with a convolution-based feature extractor pre-trained using a self-supervised objective. Whisper-medium  (789.9~M parameters) is a decoder-only multi-task architecture trained on over 680,000 hours of multilingual and multitask data using a weak supervision objective.

Each model was fine-tuned using mixed-precision training, with the AdamW optimizer \cite{loshchilov2017decoupled}, a batch size of 16 for 10 epochs, using a linear learning rate decay after a warmup over the first 10~\% of iterations. Learning rates of $1\mathrm{e}{-4}$ and $2.5\mathrm{e}{-4}$ were used for the XLSR-53 and Whisper models respectively. The XLSR-53 models were trained on a single Tesla T4 GPU with 16GB GPU memory while Whisper was trained on a RTX8000 GPU with 48GB GPU memory. In general, fine-tuning took between 24-48 hours for each model.

\section{Results and Discussion}

Benchmarking results on 19 open-source and commercial ASR systems, as well as our fine-tuning experiments, are presented in Table~\ref{tab:models_benchmarks}. 

\subsection{Large multilingual models with web-scale training data generalize better}

The overarching trend favors ASR models like Whisper \cite{radford2022robust} that were trained on vast amounts of multilingual web-scale speech data. Their data diversity and pretraining objective confer better generalization capabilities to accented speech and the clinical domain, as evidenced by their lower WER and higher Recall on MNEs, outperforming ASR models trained on monolingual data by a wide margin.

\subsection{WER vs medical WER}

As consistently observed across all pre-trained model families, M-WER was 
relatively worse overall by 4-51~\% than WER, empirically validating the performance gap on medical NEs. The only exception was GCP [Medical] where its M-WER was better,
demonstrating a trade-off in domain-specific fine-tuning.

\subsection{Relative Performance across Entity Categories}

Although Whisper-large outperformed other open- and closed-source models on WER, its MNE Recall was still poor overall with 42~\% for medications (\medication), 33~\% for protected health information (\phinfo), 59~\% for test treatment procedure (\ttp), and 67~\% for medical conditions (\condition), falling far below its practical applicability in real-world clinical scenarios \cite{luchies2018speech} due to the extent of required editing. Also, its 71~\% Recall for anatomy (\anatomy) may have resulted from the relative abundance of body parts like leg, brain, heart, liver, etc., in web-scale text.

\subsection{Medical CER: Exact vs Approximate Match}

As seen in Table~\ref{tab:medtextalign_wavlm}, medical WER sometimes unfairly penalized even the most minuscule ASR errors, e.g., quinidan vs quinidine, especially with multi-word entities, e.g., ``muscle relaxation'' vs ``mozul relaxition'', treating minor and severe ASR errors alike, a phenomenon that was not investigated in most prior works \cite{adedeji2024sound}. M-CER is complimentary in this regard, allowing us to better evaluate the severity of ASR misspellings. Also, lower M-CER helps to select the better of two ASR models with comparable WERs, like GCP~[Medical] and AWS~[Medical]. 

\subsection{Fine-tuning Results on General vs Clinical Domain}
The fine-tuned models significantly improved on MNE Recall, WER, medical WER, and medical CER, as the models were better adapted to accented speech in the clinical domain. However, this is not a silver bullet. Our results show that fine-tuning the ASR models on the general domain accented speech alone, in fact, worsens the WER, M-WER, M-CER, and Recall on clinical speech. XLSR-53 fine-tuned on the clinical subset reduced WER by 
35~\%, M-WER by 
31~\%, and M-CER by 
43~\% relative to the general domain. Fine-tuning Whisper-medium on both domains 
yielded the best results overall, improving WER by 
54~\%, M-WER by 
48~\%, and M-CER by 
65~\% relative to finetuning on the general domain only, suggesting that the ASR models still benefit from exposure to general-domain accented speech.

\section{Limitations}
ASR models, while beneficial, can risk patient safety and expose clinicians to liability through minor errors like mistranscribed drug names, doses, or diagnoses. Verification steps and spell-checkers can be integrated into the workflow to mitigate potential errors. Using automatically generated named entities instead of human annotation also introduces errors in entity identification. Automated systems can serve as initial annotation agents, with their outputs refined by domain experts. Lastly, the AfriSpeech-200 dataset often includes medical abbreviations (e.g., "Pt" for "Patient"), therefore, transcripts should be normalized for more accurate benchmarking.

\section{Conclusion}
This work highlights a noticeable disparity between general and medical WER for many SOTA ASR models, pointing to challenges in accurately recognizing accented medical named entities. Fine-tuning these models with domain-specific data was beneficial in addressing some of these issues, indicating that tailored fine-tuning can enhance ASR performance in healthcare.

\section{Acknowledgements}
We appreciate the invaluable support from Intron Health for contributing the dataset and compute for experiments.  Tejumade Afonja is partially supported by ELSA – European Lighthouse on Secure and Safe AI funded by the European Union under grant agreement No. 101070617.
We appreciate the support provided by the BioRAMP researchers, whose collaboration and insights have been fundamental to our research.

\bibliographystyle{IEEEtran}
\bibliography{main}

\end{document}